\documentclass[]{aastex701}%

\defcitealias{2024A&A...686A..42H}{H24}
\received{June 16, 2025}
\revised{July 24, 2025}
\accepted{August 7, 2025}
\submitjournal{The Astrophysical Journal}

\begin{document}

\title{A Galactic Interloper: A Study of the Cam OB1 Association's Clusters and its Visitor from the Perseus Arm}

\author[0000-0002-1650-2764,sname=Mullen,gname=Joseph]{Joseph Mullen}
\affiliation{Department of Physics and Astronomy, Vanderbilt University, VU Station 1807, Nashville, TN 37235, USA}
\affiliation{Department of Physics, University of North Florida, 1 UNF Dr., Jacksonville, FL 32224, USA}
\email[show]{joseph.mullen@vanderbilt.edu}

\author{Amanda Mast}
\affiliation{Department of Physics, University of North Florida, 1 UNF Dr., Jacksonville, FL 32224, USA}
\email{amanda.mast@unf.edu}

\author[0000-0002-5365-1267,sname=Kounkel,gname=Kounkel]{Marina Kounkel}
\affiliation{Department of Physics, University of North Florida, 1 UNF Dr., Jacksonville, FL 32224, USA}
\email{marina.kounkel@unf.edu}

\author[0000-0002-3481-9052,sname=Stassun,gname=Keivan]{Keivan Stassun}
\affiliation{Department of Physics and Astronomy, Vanderbilt University, VU Station 1807, Nashville, TN 37235, USA}
\email{keivan.stassun@vanderbilt.edu}

\author[0000-0002-1379-4204,sname=Roman-Lopes,gname=Alexandre]{Alexandre Roman-Lopes}
\affiliation{Departamento de Astronomía, Facultad de Ciencias - Universidad de La Serena, Av. Raul Bitran 1302, La Serena, Chile}
\email{aroman@userena.cl}

\author[0000-0002-3389-9142,sname=Tan,gname=Jonathan]{Jonathan Tan}
\affiliation{Dept. of Astronomy, University of Virginia, Charlottesville, VA 22904, USA}
\affiliation{Dept. of Space, Earth \& Environment, Chalmers University of Technology, Gothenburg, Sweden}
\email{jct6e@virginia.edu}

\begin{abstract}
Within the molecular clouds of the Camelopardis OB1 association exists a region previously noted as one subgroup. However, bulk clustering from Gaia astrometry has recently shown three distinctive kinematically coherent groups, all found in a similar location in the sky ($137 \lessapprox l\lessapprox145$ and $-2\lessapprox b \lessapprox5$) and at a similar distance ($\sim$1kpc). In this work, we derive from first principles the three proposed clusters in this region, refine the membership list and cluster ages, and, for the first time, examine the 3D structure, motion, and origin of the clusters. Using clustering of Gaia data in 3D position + 2D velocity space, supplemented by available SDSS-V radial velocities, we find the clusters of ages 10, 15.8, and 20 Myr with members numbering 140, 469, and 184, respectively. All three clusters overlap currently in 3D space. Tracing their previous location, based on present-day motions, shows that each cluster originated in its own distinct region and exhibited no influence on each other's formation. Two of the clusters trace their origin to different areas within the Cam OB1 association, with the oldest cluster tracing its origins to the near edge of the Perseus arm, in the direction of the Per OB1 or Cas 0B6 associations. Overall, this work illustrates how different stellar groups, even those originating in a different spiral arm, can visit and pass through each other as they travel through the Galaxy.

\end{abstract}




\section{Introduction} \label{sec:intro}
Current star formation in the Milky Way tends to occur in dense molecular cloud complexes, which can undergo gravitational collapse into clusters of stars \citep{2003ARA&A..41...57L}. Cluster stars retain the properties of their common origin and, thereby, possess similar age, chemical abundance, 3D position, and 3D motion. These clusters, which are of varying degrees of gravitational boundness, will generally survive and move together for tens, if not hundreds of Myr before being broken up and dissolving back into the disk \citep{2020SSRv..216...64K}. 


We can identify some of these newly formed clusters ($<$30 Myr) by analyzing photometric data for indicators of youth. These indicators include aspects such as infrared excess (due to the re-emission of light from circumstellar dust or a protoplanetary disk around a protostar), regions of high nebulosity (from which stars have likely recently formed), H$\alpha$ excess (as seen with T Tauri Stars), or high variability. High variability can occur due to occultation by protoplanetary disks, accretion, or strong magnetic fields (which can cause more prominent spots). These indicators of youth depend on both mass and age, and they are not without limitations and contaminants; for example, it is challenging to separate young late B, A, and F stars from field stars as they rapidly reach the main sequence. For a more detailed description of the selection of young sources/clusters, we refer the reader to \citet{2022AJ....164..137K,ABYSS} and references therein.

Another powerful technique relies on finding clusters by looking for groupings in 3D position and velocity phase space. Historically, cluster identification originated as sparse searches for overdensities when viewed in the sky \citep{2011AN....332..172S}. Distances, radial velocity (RV), or proper motion measurements could further refine cluster membership/ identification, but until recently, they were often sparsely available or imprecise. The advent of the Gaia satellite, with its unprecedented mas to $\mu as$ precision of stellar parallaxes and proper motion (depending on the magnitude of the source), has revolutionized our understanding of the structure and kinematics of the Milky Way. With accurate proper motion and parallaxes for more than 1.8 billion sources spread across the sky, Gaia data has led to many robust identifications of new clusters and cluster members (e.g., \cite{2018A&A...615A..49C,2019A&A...627A..35C,2019A&A...624A.126C,2019JKAS...52..145S,2019ApJS..245...32L,2020A&A...635A..45C,2019AJ....158..122K,2020AJ....160..279K,2024A&A...686A..42H}). Using Gaia data and hierarchical density-based spatial clustering of applications with noise (HDBSCAN; \citealt{2017JOSS....2..205M}), \cite{2020AJ....160..279K} was able to identify more than 8,000 moving groups consisting of $\sim$1 million stars. Many of these clustering techniques from phase space subsequently use the available photometry to analyze the cluster's member selection further or to derive properties such as cluster age, extinction, and distance. This is generally done through either neural networks (NN) or by directly analyzing the member's position on the HR diagram \citep{2019A&A...623A.108B,2020A&A...640A...1C,2020AJ....160..279K,2024A&A...686A..42H}.

Although Gaia marks a revolutionary advancement for identifying clusters, the analyzing works are predominantly based on 3D position and 2D proper motion measurements. Gaia provides RV measurements (i.e., the 3rd velocity dimension) for only 33 million of the 1.8 billion sources ($\sim$2\%). With the work of current and future large FOV Mulit-Object Spectroscopic Telescopes such as SDSS (Kollmeier, et al. 2025, submitted) and 4MOST \citep{2019Msngr.175....3D}, we can acquire significantly more RV measurements and further refine cluster membership. With all 6D dimensions covered, we can accurately trace the cluster back to its origin. The spectra provided can further provide information on the abundance patterns of the clusters, which can refine membership and provide hints on the nature of the molecular cloud it was formed. It is in preparation for one of these extensive spectroscopic surveys, namely the Targeting Strategy for the APOGEE and BOSS Young Star Survey (ABYSS) in SDSS-V \citep{ABYSS}, that our clusters of interest were initially highlighted (see Figure~10 therein).

ABYSS is currently observing optical (BOSS, \citealt{2013AJ....146...32S}) and near-IR (APOGEE, \citealt{2010SPIE.7735E..1CW,2017AJ....154...94M,2019PASP..131e5001W}) spectra of $\sim 10^5$ young stars across the entire sky with ages $<30$ Myr. We refer the reader to \cite{ABYSS} for more details of the observation strategy. As part of the ABYSS targeting strategy, clusters for observation were identified from the work of \cite{2020AJ....160..279K} from the clustering of Gaia DR2 data within 3 kpc of the Sun and limited to those also possessing young ages (as derived from isochrone fitting using a NN Auriga in the same work). While ensuring the RVs recovered with APOGEE and BOSS were consistent, a velocity structure was observed toward the Camelopardis OB1 association (hereafter Cam OB1) which already possesses optical and IR spectra. What is marked in prior works as one subgroup (see Section~\ref{sec:data_clustering}) appears to be three kinematically coherent groups, all found in a similar location in the sky (our Figure~\ref{fig:landb}) and at a similar distance. These groups are clearly distinguishable in proper-motion space and RV space (their Figure 10 and our Figures~\ref{fig:pm} and \ref{fig:RV}).

This work endeavors to examine the Cam OB1 association more closely and reevaluate the selection and derived cluster properties, previously done in rough en masse estimations. In Section~\ref{sec:region_overview}, we put the region of interest in the context of the wider Cam OB1 association and prior work. Section~\ref{sec:data_clustering} presents the dataset and the independently derived clustering performed. The results are presented in Section~\ref{sec:Results} where we examine in detail the derived clusters (Section~\ref{subsec:Clusters}), their ages via isochrone fitting (Section~\ref{subsec:age}), their velocity dispersion (Section~\ref{subsec:velocity_dispers}), the 3D motion and present day location of the clusters (Section~\ref{subsec:Motion}). In Section~\ref{sec:Discussion}, we will argue that one of the clusters now located in Cam OB1 was initially formed in a different spiral arm, demonstrating that coherent stellar groupings from different origins can commingle and move through or past each other in their journeys through the Galaxy.

\section{Cam OB1 Overview/ Region of Interest} \label{sec:region_overview}

\begin{figure}[!h]
    \centering
    \includegraphics[width=1\textwidth]{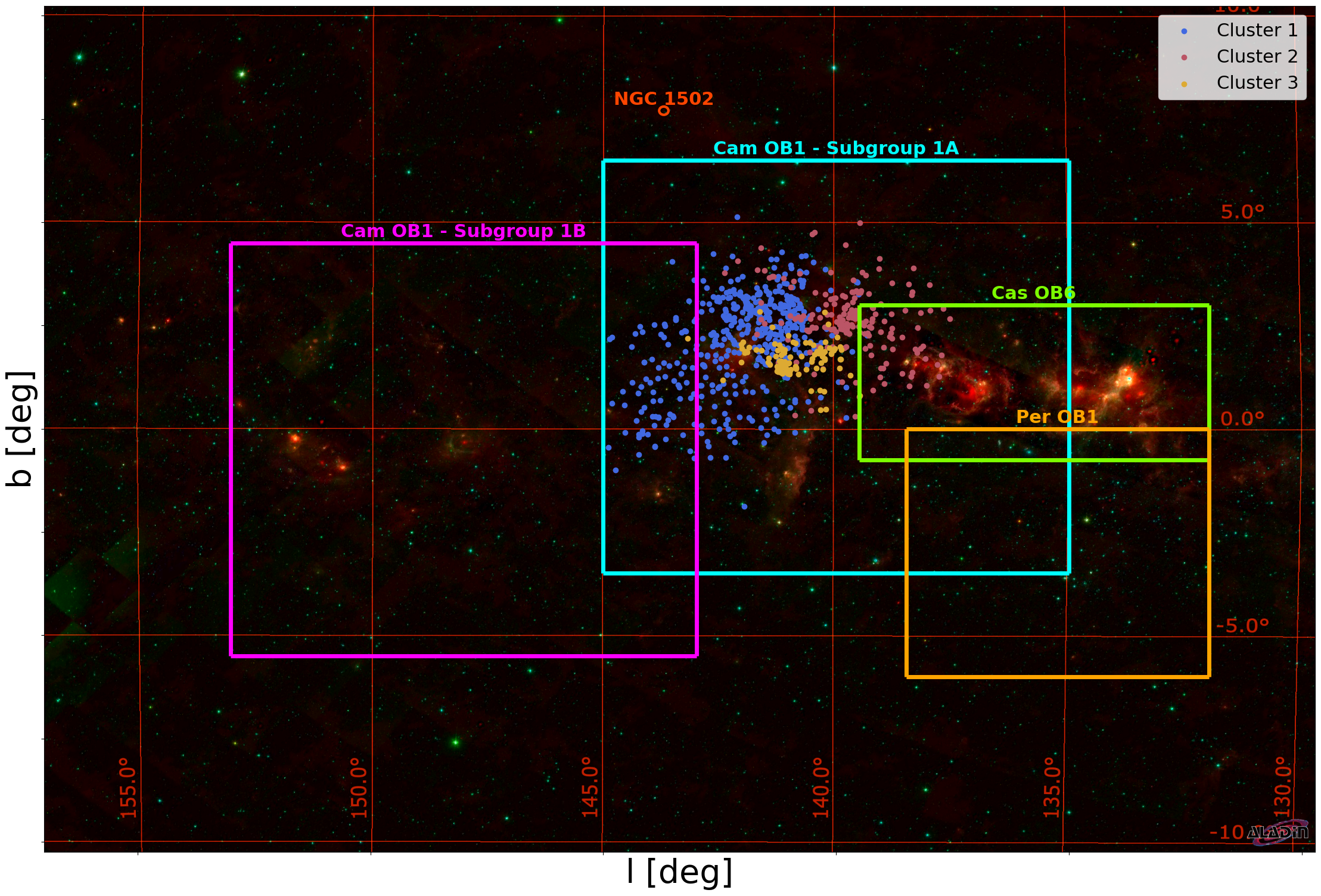}
    \caption{IR (ALLWISE) color image in the direction of Cam OB1 - Red (W4), Green (W2), Blue (W1). The four rectangles limit the areas of the associations Cas OB6 and Per OB1, located in the Perseus arm, and the generic subgroups of Cam OB1 (including NGC 1502) as listed in literature. Overplotted are the three Cam OB1 cluster members as identified by this work.}
    \label{fig:cluster_map}
\end{figure}

The region classically labeled Cam OB1 extends from $\sim 134<l<151$ and $-3<b<7$ and is characterized by its numerous dust and molecular clouds of varying densities, a plentiful YSO population and several clusters. The entirety of Cam OB1 lies around $\sim 1$kpc away at the edge of our local arm and has been proposed to have sustained star formation over the last 100 Myr \citep{2001AJ....122.2634L}. Figure~\ref{fig:cluster_map} shows an infrared (ALLWISE) image in the direction of Cam OB1, with several prominent regions noted, as discussed below.

We refer the reader to \cite{2007BaltA..16..167S}, particularly their Figure 1, for a detailed discussion of the dust clouds and nebula of this region. However, we note that the Cam OB1 molecular clouds in this region of the sky are charcterized as posessing a RV of approximately between -5 and -20km/s as measured by CO emission \citep{1996ApJ...458..561D}. The RV separates it from the nearby Gould belt layer (RV between -5 and 10km/s at $\sim$200pc) and that of the Perseus Arm (RV between -30 and -60km/s at $\sim$2-2.2 kpc), which overlap partially in projection on the sky. 

The Perseus arm associations of Perseus 0B1 (Per OB1) and Cassiopeia 0B6 associations (Cas OB6) are often noted in discussions of Cam OB1 due to their partial overlap in projection. Cas OB6 is centered on $l=134.95^\circ$, $b=0.72^\circ$ at $r=1.75$kpc with $RV=-42.6$km/s \citep{2020MNRAS.493.2339M}. It is one of the largest star-forming complexes in the Perseus arm, with extensive warm dust and HII regions extending over 150 pc along the arm (see the works of \citealt{2000ApJS..130..381C,2003ApJ...590..906T,2003ApJ...598.1005F,2006ApJ...639.1069H,2016PASA...33...46B,2020MNRAS.493.2339M} and references therein). Amongst the HII region are many young clusters such as IC 1795 with age 3-5 Myrs \citep{2012ApJ...744...87B},  IC 1805 a.k.a. the heart Nebula with age  1-3 Myrs \citep{2013A&A...554A...3S} and  IC 1848 with age 3-5Myr \citep{2014MNRAS.438.1451L}.

Per OB1 is centered on $l=134.7^\circ$, $b=-3.14^\circ$ at $r=1.83$kpc with $RV=-43.2$km/s \citep{2020MNRAS.493.2339M}. It is a well-studied region (e.g., \citealt{2008ApJ...679.1352L,2010ApJS..186..191C,2019A&A...624A..34Z,2020MNRAS.493.2339M} and references therein)  as it contains two of the richest clusters, $h$ and $\chi$ Persei, containing $>$2000 stars many of which are supergiants and O stars with ages of $\sim$13-14 Myr \citep{2002ApJ...576..880S,2019ApJ...876...65L}. Counter to Cas OB6 and Cam OB1, it does not contain an apparent giant molecular cloud nearby. Evidence suggests an expanding superbubble driven presumably by stellar winds and supernova explosions as a likely culprit \citep{2008ApJ...679.1352L,2020MNRAS.493.2339M}. The associated superbubble has been attributed to a semi-circular HI shell of size $\sim$350x500 pc centered roughly on the location of the two clusters $h$ and $\chi$ Persei \citep{1979ApJ...229..533H,2000AJ....120.1963C}. OB stars located elsewhere throughout Per OB1 are thought to have been triggered as the superbubble expanded and generally found to be younger at $\sim$5-8 Myrs old \citep{2008ApJ...679.1352L,2020MNRAS.493.2339M}.

This approximate distinction in RV between the Perseus and local arm molecular clouds has led some Cam OB1 membership papers to use available RVs to exclude potential Perseus arm member stars \citep{2007BaltA..16..167S,2013ApJ...775..123L} in lieu of accurate distance measurement. The lack of accurate individual distance measurements further necessitated membership studies to focus predominantly on easily photometrically identifiable bright young sources. Until recently, the known member stars of Cam OB1 were almost entirely composed of OB spectral types, with some later-type supergiants. Since its initial discovery by \cite{1953ApJ...118..318M}, studies such as \cite{1970A&AS....1...35H,1978ApJS...38..309H,2001AJ....122.2634L,2007BaltA..16..167S,2013ApJ...775..123L} have gradually added to the number of known OB members. 

The Cam OB1 region has previously been divided into three subgroups (1A, 1B, and 1C; see Figure~\ref{fig:cluster_map}) by \cite{2007BaltA..16..167S}, with further use of this classification by \cite{2013ApJ...775..123L}. Subgroup 1A encompasses a $\sim$10x10 deg$^2$ region, centered on (l,b)= (140.0,+1.5), with ($\mu_\alpha,\mu_\delta$)=(-1.1,-1.9) mas/yr, and 98 known members. While, subgroup 1B comprises a $\sim$10x10 deg$^2$ region, centered on (l,b)= (148.0,-0.5), with ($\mu_\alpha,\mu_\delta$)=(-0.8,-2.3) mas/yr, and 36 known members \cite{2013ApJ...775..123L}. The previously identified OB stars members in subgroup 1A are mostly concentrated around the Sh2-202 emission nebula and the vdB 14 and vdB 15 reflection nebulae. Those OB stars in subgroup 1B are primarily found surrounding a molecular cloud ring, a possible supernova bubble remnant that has left the region inside lacking gas and dust to form new stars. Additional OB stars in 1A and 1B are located along the filamentary structure of the cloud complex (see Figure 1 of \cite{2007BaltA..16..167S}). Isochone fitting to subgroup 1A and 1B OB members returns similar ages of at least 10–15 Myr, with some possible members suggesting a larger age distribution \citep{2013ApJ...775..123L}.

Subgroup 1C consists primarily of the cluster NGC 1502 \citep{1991MNRAS.253..649T,1997AJ....114..222W,2005A&A...443..157P} and therefore has a well-constrained age ($\sim$10Myr) and distance of (1180$\pm$160pc). It occupies  0.1 deg centered on (l,b)=(143.7,+7.7), with ($\mu_\alpha,\mu_\delta$)=(-0.2,0.4) mas/yr, and has 76 known members. Recent searches for overdensities \citep{2010AstL...36...75G,2013ApJ...775..123L} have identified a new cluster designated G144.9+0.4. This cluster is located in between subgroup 1A and 1B at (l,b)=(144.9,0.4), is $\sim$7' across (2pc across), and has 23 candidate members. The presence of nebulosity and classical T Tauri stars suggests a young age of 1-2Myr, corresponding to one of the latest episodes of star fomation in Cam OB1. 

The groups above share similar proper motions and ages, point to a similar formation history within the Cam OB1 region. However, the region's sizeable physical extent of $\sim$320x160pc$^2$  on the plane of the sky \citep{2013ApJ...775..123L} implies a coeval evolution to be impossible \citep{1999AJ....117..354D} within the association. Furthermore, all these discovering works are based upon early-type stars representing the youngest, shortest-lived members, which may not indicate all the sub-structures within the group. In this work, we focus on the recoverability and characterization of the three kinematic structures initially proposed in \cite{2020AJ....160..279K} and as seen in Figure 10 in \cite{ABYSS}, or this works Figures \ref{fig:landb}, \ref{fig:pm}, and \ref{fig:RV}. Note that all three substructures in this work fit solely within the on-sky region denoted as Subgroup 1A, with minimal overlap with subgroups 1B or 1C and only one star in the vicinity of G144.9+0.4. Subgroup 1A has the widest velocity dispersion of any of the aforementioned literature subgroups ($\sim \pm$ 4mas/yr), and the proper motions of the stars in this work fit well within the range.

The work presented in the following sections is unique compared to other Cam OB1 studies. We use all available Gaia sources and, therefore, make no selection in spectral type (i.e., we do not limit ourselves to OB spectral types). With more sources and more accurate proper motions, we are able to see additional substructure. Most importantly, unlike other works, Gaia provides accurate distance measurements, and therefore, we do not use any cuts in RV as an initial filtering for Cam OB1 stars.

\section{Data and Methods}\label{sec:data_clustering}

This work predominantly uses photometry, proper motion, and parallax from Gaia DR3 \citep{2021A&A...649A...1G}. We further supplement RV measurements from SDSS-V \citep{2009ApJ...705.1320G} when available and applicable. To test the recoverability of the three Cam OB1 clusters as initially derived in \cite{2020AJ....160..279K} and noted in \cite{ABYSS}, very generous selections were made on position, proper motion, and magnitude, namely:
\begin{itemize}
    \item $\alpha,\delta$: between $\pm10^\circ$ centered on $\delta=59^\circ$, $\alpha=$3h30m
    \item $\mu_\delta$ and $\mu_\alpha$ between $\pm3$ mas/yr 
    \item $\varpi$ between $0.7-1.1$ mas
    \item $G<18$ mag
\end{itemize}

To isolate potential clusters in this region, we use a Python implementation of the clustering algorithm "hierarchical density-based spatial clustering of applications with noise" (HDBSCAN; \citealt{2017JOSS....2..205M}). HDBSCAN has been extensively utilized in blind searches for star clusters, not limited to but including the works of \citep{2022A&A...659A..59T,2019AJ....158..122K,2020AJ....160..279K,2023A&A...673A.114H,2024A&A...686A..42H}, where these clusters were initially isolated. HDBSCAN works by condensing the minimum spanning tree by pruning off the nodes that do not meet the minimum number of sources in a cluster and reanalyzing the nodes that do. Depending on the chosen algorithm, it would then either find the most persistent structure (through the excess of mass method) or return clusters as the leaves of the tree (which results in somewhat more homogeneous clusters, \citealt{2019AJ....158..122K}).

The two main parameters that control HDBSCAN are the number of sources in a cluster and the number of samples. The former rejects too small groupings; the latter sets the threshold for how conservative the algorithm is in considering background noise (even if the resulting noisy groupings meet the minimum cluster size).

Similar to \cite{2020AJ....160..279K}, the clustering was performed on the 5D data set: Galactic coordinates $l$ and $b$, parallax, and proper motions. The conversion from the equatorial to the Galactic reference frame for the positions themselves is necessary, as most of the structure is located along the Galactic plane, and the cos $\delta$ term would add nonlinear distortions in $\alpha$ otherwise. In terms of proper motion, we choose to cluster in $V_l$, $V_b$ instead of $V_\alpha$, $V_\delta$; however,  these are direct rotational transformations of each other, and thus they produce broadly comparable outputs. Proper motions were converted to the local standard of rest (LSR) reference frame (using constants from \cite{2010MNRAS.403.1829S}) to avoid distortions due to the line of sight from the solar motion, as well as converted to the physical units of km/s to prevent the distortion in the distance. Converting proper motions to velocities is generally a nontrivial issue, as converting Gaia parallaxes to distances can result in asymmetric errors and non-Gaussian parameter distributions \citep{2021AJ....161..147B}. \cite{2024A&A...686A..42H} (hereafter H24) noted that this conversion had no impact on nearby clusters' detectability and membership lists, and the main benefits started for clusters more distant than $\sim$two kpcs (twice the distance of Cam OB1). Effects, though, maybe more prevalent at closer distances for more diffuse moving groups and extended structures than those studied in \citetalias{2024A&A...686A..42H}. 

Various scaling factors were considered to normalize each of the five dimensions. We noted that parallax and/or distance played the key role in the recoverability of the clusters and found that clustering in log(parallax) provided optimal results; everything else was left in its native units (i.e., degrees, mas, and km/s). In this work, we required the minimum number of samples to be 24 sources and the minimum number of stars per cluster to be 15 stars. The leaf method was utilized as it could better separate distinct clusters from other sources in the field. In contrast, the excess of mass (eom) method tended to overly merge the clusters, as similarly noted in \cite{2019AJ....158..122K}.

Although powerful, we should note that HDBSCAN is prone to oversensitivity and can easily create false positive clusters due to mere statistical fluctuations and not real astrometric overdensities. To mitigate this issue, we ran all potential clusters through an isochrone fitting NN, Auriga \citep{2020AJ....160..279K}, to get a rough approximation of their age and eliminated those clusters older than $10^{7.5}$ yrs old ($\sim$31.5 Myrs). This age constraint is consistent with the age of the target young clusters found in \citep{2020AJ....160..279K}, and what can be expected from clusters we initially assumed formed in Cam OB1.

\section{Results}\label{sec:Results}
\subsection{The Clusters}\label{subsec:Clusters}
\begin{figure}[!h]
    \centering
    \includegraphics{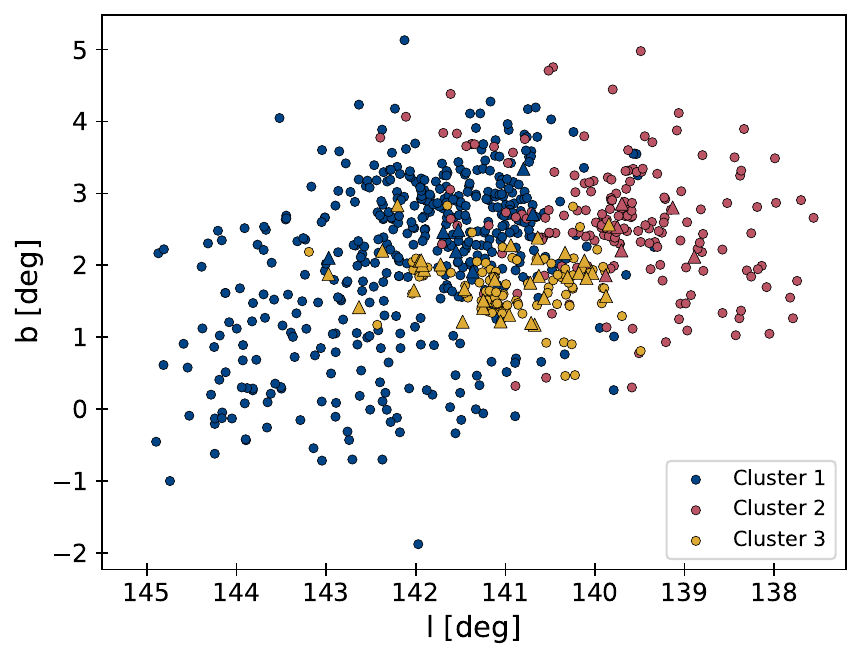}
    \caption{The spatial distribution of members of the three Cam OB1 clusters as identified by this work's clustering algorithm. Circles denote sources found in the initial clustering algorithm, while triangles represent sources manually added due to the availability of additional RV measurements.}
    \label{fig:landb}
\end{figure}

\begin{figure}[!t]
    \centering
    \includegraphics{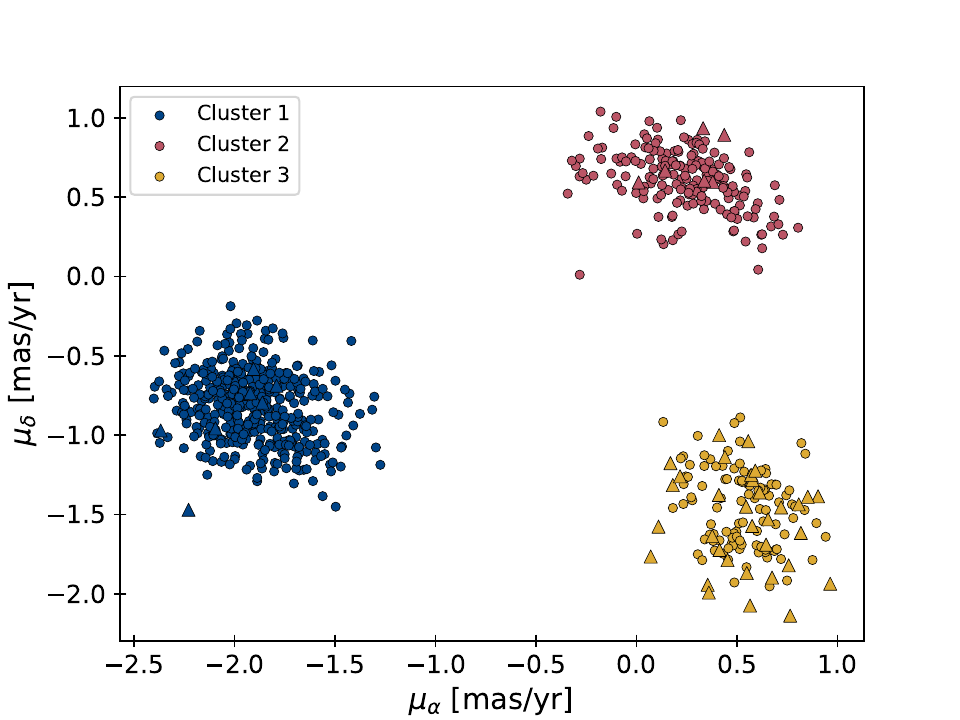}
    \caption{Velocity structure of cluster members observed toward the region of the Cam OB1 association shown in Figure~\ref{fig:landb} and as identified by this work's clustering algorithm. Each cluster occupies a distinguishable location in proper motion space where colors/ symbols are consistent with those of Figure~\ref{fig:landb}.
}
    \label{fig:pm}
\end{figure}

\begin{figure}[h]
    \centering
    \includegraphics[width=0.75\textwidth]{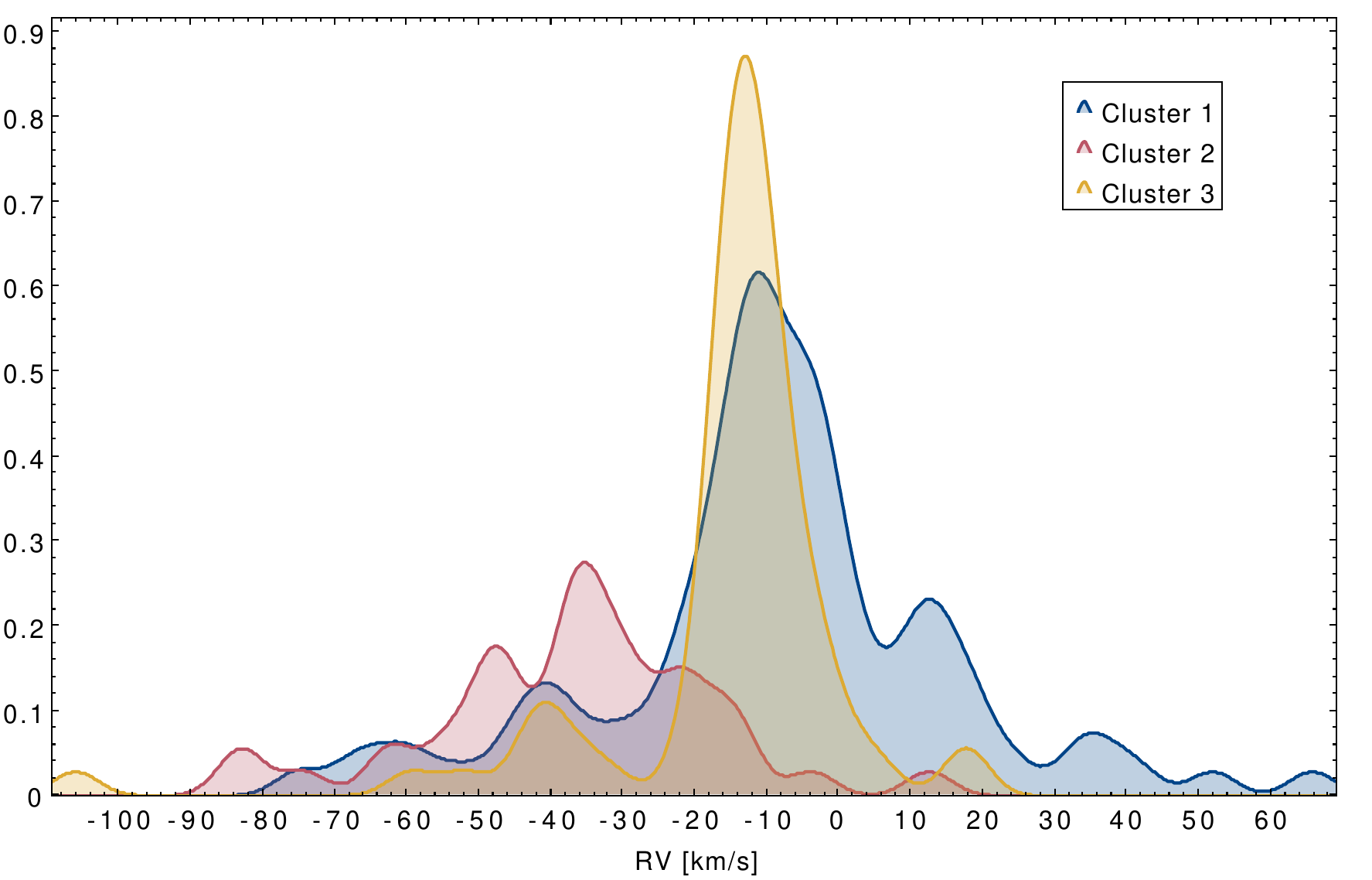}
    \caption{Radial velocity distributions for the clusters shown in Figures~\ref{fig:landb} and ~\ref{fig:pm} and their subset of sources with available RV measurements.  }
    \label{fig:RV}
\end{figure}

With these minimal constraints, we independently recovered the three clusters in question. We also searched for additional substructures that previous works might have missed with a more finely-tuned approach, but none were found. Table~\ref{tab:cluster_glob} shows the global properties of each cluster. The clusters are predominantly located in the same region of the sky as shown in Figure~\ref{fig:landb}, but have distinct proper motion distributions, shown in Figure~\ref{fig:pm}. Although not part of the clustering algorithm, Figure~\ref{fig:RV} plots the subset of cluster sources with available SDSS RVs.

To further enhance our cluster candidate, we add back in previously classified field sources who have RVs available where the source's RV falls near the peak of the cluster's RV distribution (Cluster 1: $-30 <RV< 25$ km/s; Cluster 2: $ -55<RV<-15$ km/s; Cluster 3: $ -20<RV<0$ km/s) and lies within the distribution of proper motion, $l$, $b$, and parallax consistent to each cluster. These additional sources (7 for Cluster 1, 6 for Cluster 2, and 33 for Cluster 3) are shown as triangles in Figures~\ref{fig:landb} and \ref{fig:pm}. We note here that the RVs of Cluster 1 and 3 members approximately correspond to what we see from molecular clouds corresponding to Cam OB1 and, likewise, what we expect of its constituent stars (see Section~\ref{sec:region_overview}). On the other hand, the members of Cluster 2 have more 
systematically blue-shifted RVs by $\gtrapprox$ 21 km/s. Prior works on Cam OB1 would have a priori excluded these sources as their RV values are more similar to those found within molecular clouds of the Perseus arm. However, with our accurate astrometry (see Section~\ref{sec:Results}), we can confidently place Cluster 2 within the same 3D region of space associated with Cam OB1.

The individual member candidates for each cluster are listed in Table~\ref{tab:cluster_indiv}. It is worth mentioning that individual member candidates for each cluster vary from those of the similar clusters found in \cite{2020AJ....160..279K} and \citetalias{2024A&A...686A..42H} (in terms of the additional source added/removed, not cluster labeling). Still, the bulk properties of each cluster (i.e., age, position, motion) remain similar. These discrepancies are a natural consequence of different implementations of HDBSCAN with slight differences in minimum sample size, minimum cluster size, the exact form of the 5Ds clustered in, and any further validation cuts (see each respective paper for details). 

For instance, some noticeable differences include \citetalias{2024A&A...686A..42H} generally having more sources in the interior of the distribution, which can be attributed to setting a higher threshold on the minimum cluster size. \citetalias{2024A&A...686A..42H} clusters also generally have fewer sources in the physical periphery due to post-clustering validation where potential cluster members farther from the bulk of the cluster were discarded. In comparison to our clusters (Figure 9, right), the total cluster size defined in \citetalias{2024A&A...686A..42H} is much smaller ($R_{Total}$: Cluster 1=47pc; Cluster 2=22pc, Cluster 3=18pc). Although this is an essential step in bulk OC searches to ensure reliable cluster parameters are derived, it most likely does not represent the actual distribution of members and removes any potential cluster tails. 

Within the same smaller on-sky region as the clusters defined in \citetalias{2024A&A...686A..42H}, we recover ~2/3 of their members for each of our clusters. There are additional potential sources (Cluster 1=121; Cluster 2=45; Cluster 3=73) listed in \citetalias{2024A&A...686A..42H}'s catalog that are not included in ours. Approximately half of our sample is not in \citetalias{2024A&A...686A..42H}'s sample, due to either being outside the regions defined in \citetalias{2024A&A...686A..42H} (Cluster 1=35\%; Cluster 2=43\%; Cluster 3=36\%), or occupying the same on-sky region as \citetalias{2024A&A...686A..42H} but only found in our sample (Cluster 1=15\%; Cluster 2=6\%; Cluster 3=13\%). Lastly, it is worth noting that \citetalias{2024A&A...686A..42H} divides our Cluster 3 into two parts. However, we see no evidence of any reason for subdivision in phase space and choose to keep Cluster 3 as a whole in our work/comparison.

\begin{table}
\caption{Global Subgroup Properties\label{tab:cluster_glob}}
\begin{center}
\begin{tabular}{lcccccccc}
\tableline
\colhead{Subgroup} & \colhead{(l,b)} & \colhead{Angular Size} & \colhead{Distance} & \colhead{Age} & \colhead{Num. Stars} & \colhead{$\mu_{\alpha}$} & \colhead{$\mu_{\delta}$} & \colhead{RV}\\
\colhead{} & \colhead{(deg)} & \colhead{(deg)} & \colhead{(pc)} & \colhead{(Myr)} & \colhead{} & \colhead{(mas yr$^{-1}$)} & \colhead{(mas yr$^{-1}$)} & \colhead{(km/s)}\\
\tableline
\tableline
Cluster 1 & 142.00, +2.11 & $\sim$5.6 x 7.0 & 1095 $\pm$ 43 & 15.8 $\pm$ 1.9& 469 & -1.91 $\pm$ 0.21 & -0.82 $\pm$ 0.21 & -8.9 $\pm$ 24.4 \\
Cluster 2 & 139.72, +2.59 & $\sim$4.8 x 4.7 & 1064 $\pm$ 26 & 20.0 $\pm$ 2.4 & 184 & 0.24 $\pm$ 0.24 & 0.60 $\pm$ 0.19 & -37.1 $\pm$ 19.5\\
Cluster 3 & 140.96, +1.72 & $\sim$3.7 x 2.4 & 1090 $\pm$ 43 & 10.0 $\pm $1.2 & 140 & 0.53 $\pm$ 0.18 & -1.48 $\pm$ 0.26 & -15.7 $\pm$ 17.4\\

\tableline
\end{tabular}  
\end{center}

\end{table}

\begin{table}
\caption{Individual Member Candidates\label{tab:cluster_indiv}}
\begin{center}
\begin{tabular}{lcccccccc}
\tableline
\colhead{Gaia ID} & \colhead{(l,b)} & \colhead{$\mu_{\alpha}$} & \colhead{$\mu_{\delta}$} & \colhead{RV}  & \colhead{$J$} & \colhead{$H$} & \colhead{$K$} & \colhead{Cluster}\\
\colhead{} & \colhead{(deg)} & \colhead{(mas yr$^{-1}$)} & \colhead{(mas yr$^{-1}$)} & \colhead{(km/s)} &\colhead{(mag)} & \colhead{(mag)} & \colhead{(mag)} & \colhead{}\\
\tableline
\tableline
462947141792608256	& 139.643, 2.475 & 0.35$\pm$0.02 &	0.64$\pm$0.02 & \nodata	&	12.46 &	12.02 &	11.91 &	Cluster 2\\
462934497409493248	& 139.900, 2.423	& 0.30$\pm$0.01	& 0.86$\pm$0.02	& \nodata	& 9.54	& 9.45 &	9.34 &	Cluster 2\\
462930065003274368	& 140.005, 2.347	& 0.60$\pm$0.02	& 0.46$\pm$0.02	& -35.02	& 12.96 &	12.61	& 12.46 &	Cluster 2\\
462930305521422976	& 140.057, 2.416	& 0.11$\pm$0.07	& 0.37$\pm$0.07	& \nodata	& 14.39 &	13.68	& 13.44	& Cluster 2\\
462180816549810816	& 140.105, 2.401	& 0.55$\pm$0.03	& 0.64$\pm$0.03	& \nodata	& 13.24 &	12.92	& 12.77 & 	Cluster 2\\
462924120768568064	& 139.975, 2.218	& 0.30$\pm$0.02	& -1.01$\pm$0.02	& \nodata	& 12.48 &	11.75	& 11.03	& Cluster 3\\

\tableline
\end{tabular}  
\end{center}
\tablecomments{Table~\ref{tab:cluster_indiv} is published in its entirety in the machine-readable format. A portion is shown here for guidance regarding its form and content.}
\end{table}

\subsection{Age} \label{subsec:age}

To determine the age of each cluster, we employed traditional isochrone fitting techniques. While neural networks (NNs), such as Auriga (see Section~\ref{sec:data_clustering}) and those utilized by \cite{2020A&A...640A...1C,2020AJ....160..279K,2024A&A...686A..42H}, are powerful for approximating parameters for large catalogs of clusters, their training on potentially sparse or simulated datasets can lead to systematic uncertainties, particularly for individual cases \citep{2020AJ....160..279K}. Therefore, for the detailed study of individual clusters, this work utilized visual isochrone fitting based on the entire color-magnitude diagram. This approach allows for a more direct accounting of observational nuances such as outliers and unresolved blended/binary sources, which is crucial for refining age estimates. For a detailed review of the strengths and limitations of isochrone fitting for open clusters, we refer the reader to works such as \citet{1994ApJS...90...31P}.

\begin{figure}[!h]
    \centering
    \includegraphics[width=0.5\textwidth]{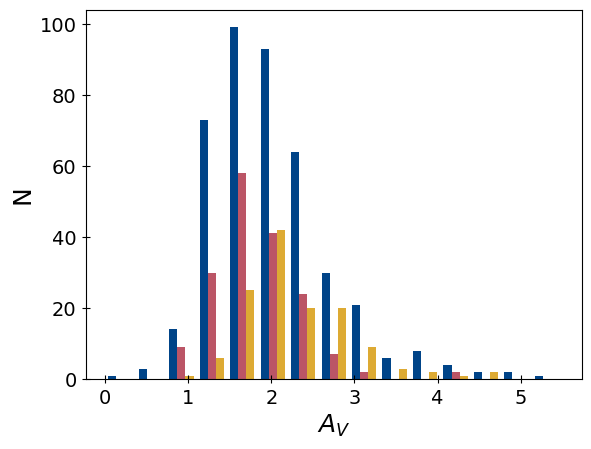}
    \caption{Extinction distribution for the three Cam OB1 clusters.}
    \label{fig:AV}
\end{figure}

In this work, we fit directly MIST isochrones \citep{2016ApJ...823..102C} to the absolute $G$ magnitude vs. $BP-RP$ color-magnitude diagram. Distances to individual stars were calculated via direct inversion of their Gaia parallaxes ($d = 1/\varpi$). While this method can be subject to biases for individual stars, especially at larger distances or for lower precision parallaxes (as discussed in Section~\ref{sec:data_clustering}, \citealt{2021AJ....161..147B}), its impact is generally reduced for clusters within $\sim$1 kpc like those studied here \citep{2024A&A...686A..42H}.  We downloaded photometry and Gaia XP spectrum (if it was available). Using Teff and log g obtained from SDSS spectra \citep{2024AJ....167..173S} or Gaia Net \citep{2025arXiv250302958H}, we picked the best matching synthetic BT-settl spectrum. We then performed least squares fitting using SEDFit to measure Av using \cite{2009ApJ...705.1320G} profile. The distribution of extinction for each cluster is shown in Figure~\ref{fig:AV}. 

\begin{figure}[!h]
    \centering
    \includegraphics[width=1.0\textwidth]{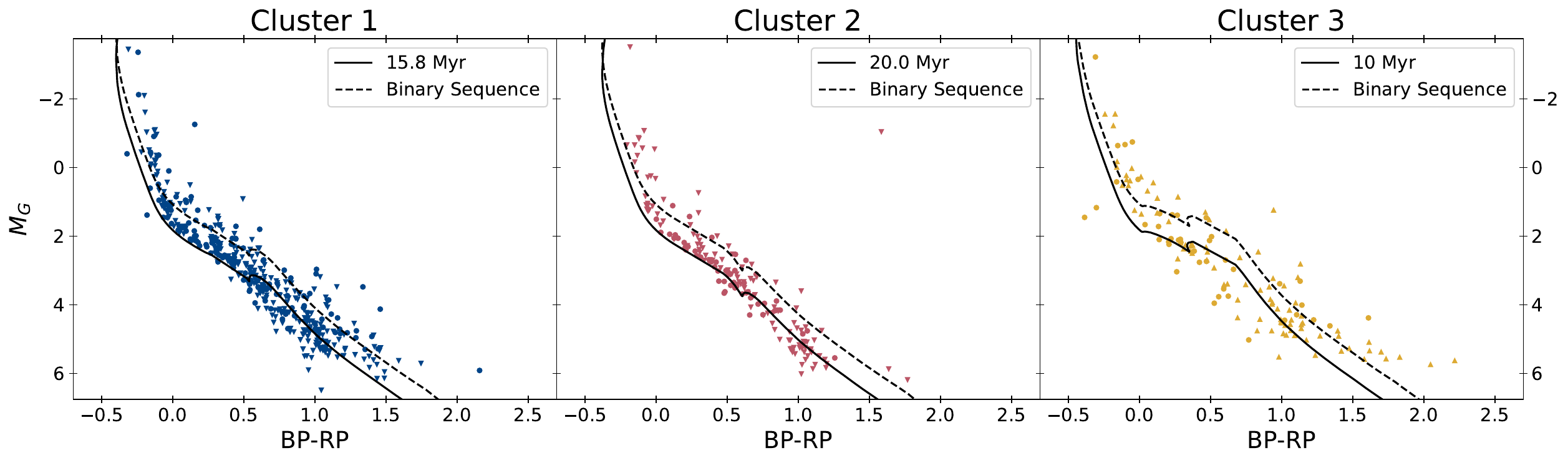}
    \caption{Color-magnitude diagrams for the three Cam OB1 cluster members, where absolute G-mag and BP-RP colors have been corrected for interstellar extinction using extinctions calculated from $T_{eff}$ and logg from SDSS (dots) or Gaia (triangle). The solid delineates the best-fit MIST isochrone, whereas the dotted line corresponds to the same age binary sequence (i.e., doubling the brightness in linear space).}
    \label{fig:isochrones}
\end{figure}

Conversions from V band extinction to G, BP, and RP band values come directly from the PARSEC models \citep{2012MNRAS.427..127B}, which adopts the extinction law from \cite{1989ApJ...345..245C} and integrates synthetic ATLAS9 spectra with the Gaia nominal passbands \citep{2010A&A...523A..48J}. The best matching isochrones are plotted in Figure~\ref{fig:isochrones}, where the dotted line corresponds to the same age binary sequence (i.e., doubling the brightness in linear space).

Cluster 3 appears the youngest at 10$\pm$1.2 Myr old, followed by Cluster 1 at 15.8$\pm$1.9 Myr and Cluster 2 at 20.0$\pm$2.4 Myr.The quoted age uncertainties are based solely on the grid spacing of the MIST isochrones, and the errors are thus lower limits, as they do not fully account for additional photometric errors or systematic uncertainties. Clusters 1 and 2 fit the isochrones well, while there is a slightly larger dispersion around the best fit for cluster 3. The young age of Cluster 3 can contribute to the slightly smaller velocity dispersion compared to that of the older Cluster 1, as seen in 1D in Figure~\ref{fig:RV}.

These ages follow the same relative trend as the NN of \citetalias{2024A&A...686A..42H}, which showed the analogous average age of Cluster 3 at 4.15 and 4.85 Myr (labeled as two separate clusters, see Section~\ref{sec:data_clustering} discussion), Cluster 1 at 14.60 Myr, and Cluster 2 at 34.66 Myr. The discrepancy between the ages in this work and that of \citetalias{2024A&A...686A..42H} serves to illuminate the inherent uncertainties amongst isochrone fitting techniques. If one were to err on the side of more extreme age differences between the clusters, as in \citetalias{2024A&A...686A..42H}, it would only strengthen the arguments being made in Sections~\ref{subsec:Motion} and \ref{sec:Results} (with the youngest Cluster 3 having less time to move and the oldest Cluster 2 having even more time to travel). Our derived ages incidentally represent a more moderate approach to the 3D motion. In the following sections, we will show how the older age of Cluster 2 mixed with its 3D motion provides a unique insight into its origin.


\subsection{Velocity Dispersion} \label{subsec:velocity_dispers}
We note that the clusters recovered have generally larger velocity dispersions than those typically associated with open clusters. The clusters posses the overall mean motion of Cluster 1: $( 20.7 \pm 14.0 , 12.5 \pm 11.2 , -2.4 \pm 1.3 )$ with a velocity dispersion of $\sigma=18.0$ km/s; Cluster 2: $( 38.6 \pm 10.3 , -10.4 \pm 9.0 , 9.1 \pm 0.9 )$ with $\sigma=13.7$ km/s; and Cluster 3: $( 17.4 \pm 9.6 , -1.3 \pm 7.5 , 1.7 \pm 1.2)$ with $\sigma=12.3$ km/s. In comparison, in the 30 Myr NGC 2547 near Vel OB2, the velocity dispersion is on the order of 5 km/s \citep{2018MNRAS.481L..11B}, and it is usually much less than that. The only other well-studied region with a more extreme scatter is Ophion \citep{2025arXiv250302958H}, which is entirely incoherent.

{The larger velocity dispersion in Cam OB1 clusters is strongly seen in RVs (Figure~\ref{fig:RV}) and less so in the PMs (Figure~\ref{fig:pm}). This difference is likely due to the clustering algorithm utilizing proper motion, which may reject diffuse members with too discrepant PMs. Contrarily, we did not cluster in RV due to a lack of available RVs, and some discrepant RV sources may be included in our sample. Most likely, the true velocity dispersion lies somewhere between those seen on the plane of the sky and those seen in the RVs. 


To further analyze if potential contamination is artificially inflating the velocity dispersion, we note that each cluster has a RV error distribution with a median value of $\sim$2.4km/s and a long-tail of higher RV errors (typical of APOGEE RVs), with 21\% of the data having RV $>10$km/s (primarily high-mass stars). Additionally, our sample includes two candidate spectroscopic binaries (Gaia Source ID: 461821551124212992 and 461805264608251008), which are characterized by a positive SDSS starflag (i.e., ``$MULTIPLE\_SUSPECT$"; denoting a double-peaked cross correlation function) and that also possess a high scatter of visit-level RVs around the SNR-weighted average RV (``vscatter" = 23km/s, 35km/s respectively). There are likely other spectroscopic binaries in our sample that either do not have RVs at all or with sufficient epochs ($\ge 3$ ) to confirm binarity; however, we do not expect binary contamination to exceed 5\% of our total sample. If those with larger RVs ($\gtrapprox$ 2.4km/s) or potential spectroscopic binaries were to be excluded, the bulk average motion plotted and discussed in Section~\ref{subsec:Motion} remains overall unaffected, with changes in velocity dispersion of less than $\sim$1km/s.

Upon investigation, we did not note any particularly strong substructures in proper motion throughout the clusters, with the exception of a slight signature of outward expansion as seen in Clusters 1 and 2. Figure~\ref{fig:substructure} shows the stellar motion in the plane of the sky relative to the average motion of each individual cluster, in both direction and magnitude for each cluster.

\begin{figure}[!h]
    \centering
    \includegraphics[width=1\textwidth]{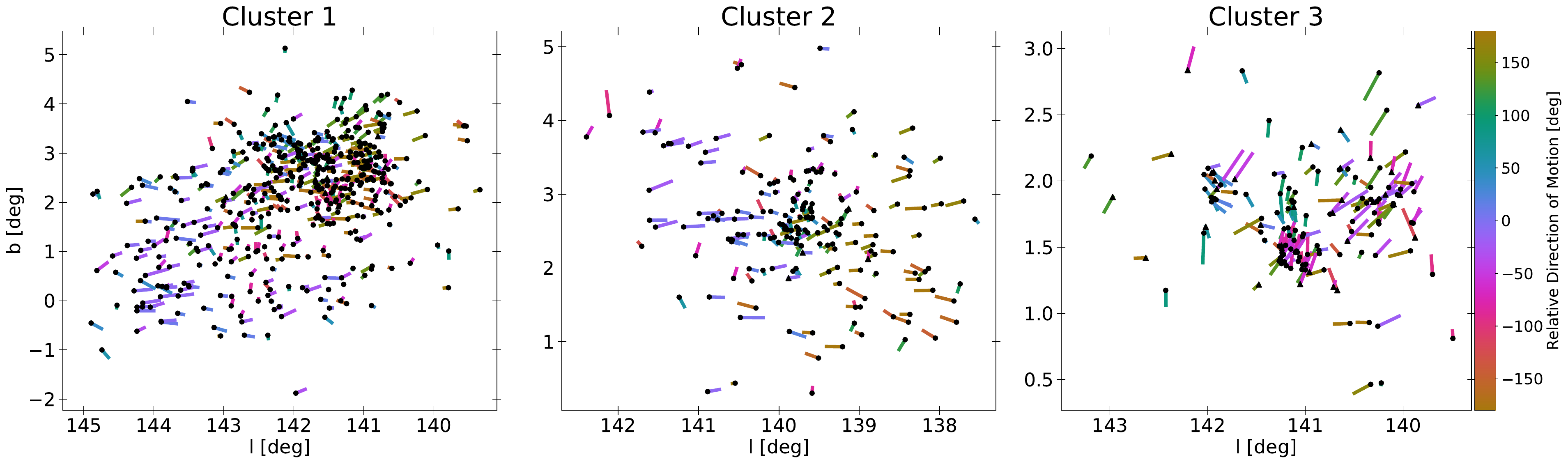}
    \caption{Location of member stars (black points) and their residual proper motion vectors after subtracting the mean motion of their respective cluster. Motion vectors are directed toward the present-day location of the star with colors based on the direction of motion (i.e., Clusters 1 and 2 show a signature of expansion).}
    \label{fig:substructure}
\end{figure}

\subsection{Motion \& Present Day Location} \label{subsec:Motion}
In this section, we trace each star's motion over time by utilizing the subset of cluster members with RV measurements. Figure~\ref{fig:xyz} traces back the median position of each cluster in galactic coordinates from their current location (solid dot) to its approximate origin corresponding to the age derived in this work. Figure~\ref{fig:proximity} traces the relative distance between the mean position of each cluster and also the current relative size of each cluster.

\begin{figure}[!h]
    \centering
    \includegraphics[width=0.75\textwidth]{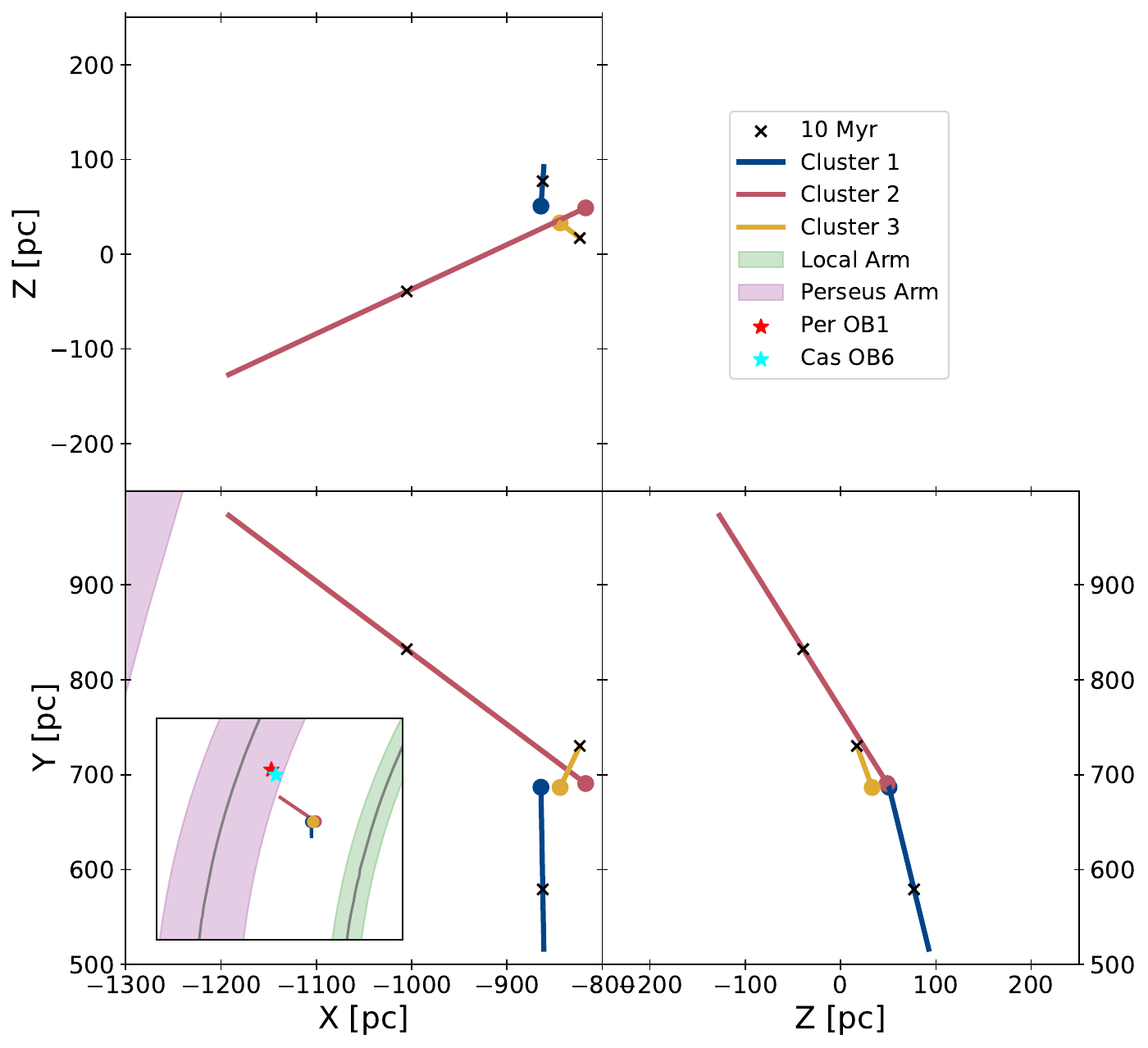}
    \caption{The motion of the three clusters from their formation location to their current position (large dot). To help show the clusters' motion with their varying ages, an ``X" has been placed along each path corresponding to each cluster's position 10 Myr ago (i.e., the age of the youngest cluster). The bottom-left panel includes a zoomed-out inset version of the same panel, with the shaded regions corresponding to the approximate current location of the Perseus and Local Arm as defined by \cite{2019ApJ...885..131R}. The central location of the Per OB1 and Cas OB6 associations \citep{2020MNRAS.493.2339M} is marked with a star.}
    \label{fig:xyz}
\end{figure}

\begin{figure}[!h]
    \centering
    \includegraphics[width=\textwidth]{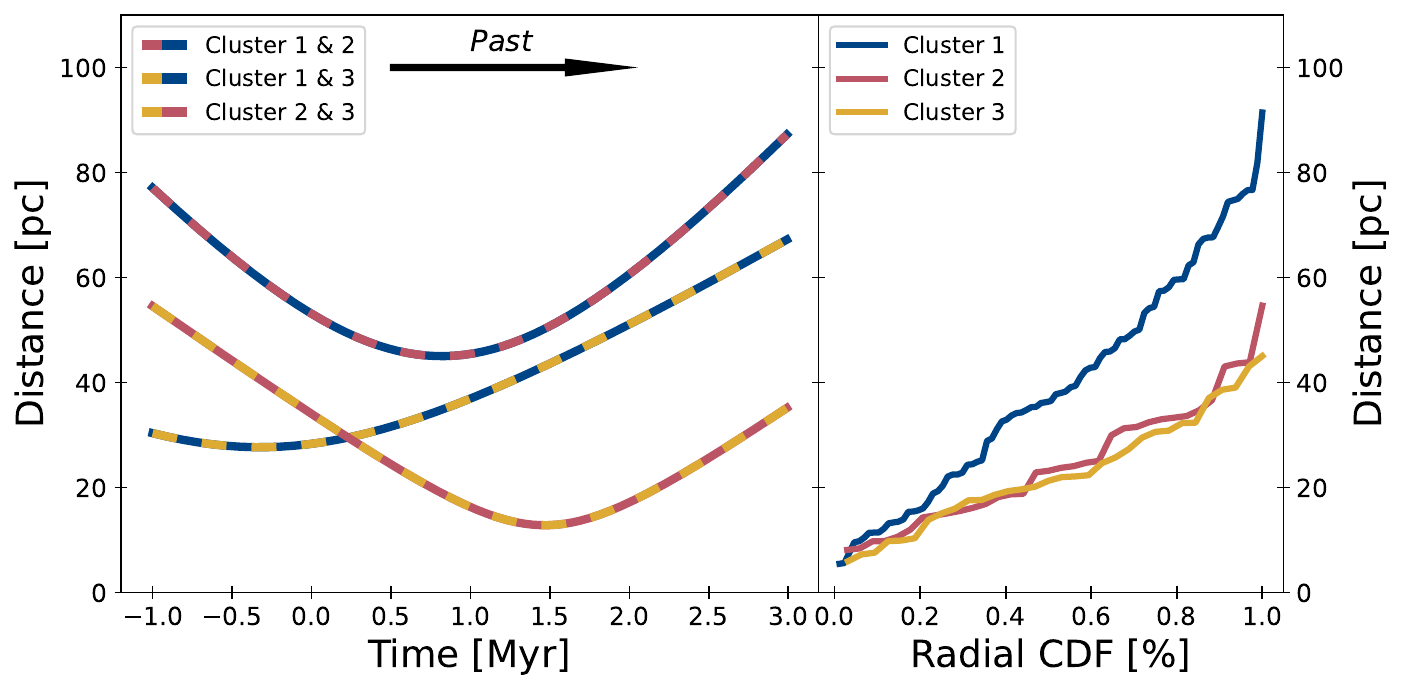}
    \caption{\textit{\textbf{Left:}} The relative distance between the average center of each cluster with time. The plot extends from the future (neg x-values), where clusters 1 and 3 will approach their closest relative position, to the near past (positive x-values). The trends to the right of this plot extend linearly to more past times until each cluster reaches its respective formation place/time (see Figure~\ref{fig:xyz}). \textit{\textbf{Right:}} The cumulative percent of total cluster sources within a given radius from each cluster's central location at the present time.}
    \label{fig:proximity}
\end{figure}

We assume the sources retained their present-day motions and RVs. The simplification of constant cluster velocity does not consider external accelerating influences, such as Cluster 2 most likely crossing through the galactic plane. However, we can clearly rule that these clusters have formed independently of one another and are happening to coincide at this approximate moment.

Figures ~\ref{fig:xyz} and ~\ref{fig:proximity} show that each cluster approaches its closest relative position in their near past, much shorter than the age of the respective clusters (i.e., comparing Cluster 2 to Cluster 3 $\sim$1.5 Myr ago or to Cluster 1 $\sim$0.8 Myr ago). Cluster 1 and 3 will be at their closest position sometime between the present and $\sim$0.3 Myr from now. We note that at the current time, all clusters partially overlap in 3-dimensional space. The radius from the clusters' central location to that containing 50\% of each cluster's members is: Cluster 1 $R_{50}$=36.5 pc; Cluster 2 $R_{50}$=23.5 pc; Cluster 3 $R_{50}$=21.6 pc. The convergence shown in Cam OB1 contrasts with other star-forming regions, such as Orion, where everything expands ballistically, and the stars would have been closest at the time of formation \citep{2018AJ....156...84K}.

At the present time, we cannot attribute the current convergence to anything more than a chance encounter due to their own respective motions. In the past, classical OB associations often had to be subdivided into multiple regions (e.g., in Cepheus and Cass), even some along the same line of sight. However, they were usually separated in their distance, and it would not be unexpected to find multiple unrelated populations at different distances along the Galactic plane. There are a few examples of multiple populations inhabiting similar volumes of space (e.g., Vel OB2 \& NGC 2547—though these formed there to begin with, they didn't migrate from elsewhere, and one has likely caused the formation of another - \citealt{{2019A&A...621A.115C}}). Two populations clearly coming from different regions would have already been peculiar, but three of them passing by each other so closely at the same time is beyond unexpected.

\section{Discussion: Origin of Clusters and Interlopers}\label{sec:Discussion}
Figure ~\ref{fig:xyz} shows that Cluster 1 and 3 formed within the confines of Cam OB1, having only traveled $\sim$240 pc and $\sim$65pc from their respective formation location. No influence should be attributed for one cluster triggering the formation of another cluster. At the time of Cluster 3's formation (marked by X's in Figure~\ref{fig:xyz} at 10Myr ago for each cluster), Cluster 1 was over $\sim$170 pc away with Cluster 2 farther. Similarly, Cluster 2 did not trigger formation in Cluster 1.

Cluster 2 presents a unique case among the three. At $\sim$20 Myr, it is the oldest, and several lines of evidence point to its origin within the Perseus Arm, distinguishing it from Clusters 1 and 3 which originated within Cam OB1. Its overall 3D motion is significantly faster and incongruous with the other two clusters (Figure~\ref{fig:xyz}), having traversed over 550 pc since its formation. Furthermore, its radial velocity distribution (Figure~\ref{fig:RV}, Table~\ref{tab:cluster_glob}) is systematically blue-shifted compared to the other clusters and is more consistent with the kinematics of structures within the Perseus Arm rather than the Local Arm (see Section~\ref{sec:region_overview}). Indeed, projecting Cluster 2's motion backward places its formation locus near the edge of the Perseus Arm, as defined by \cite{2019ApJ...885..131R} and illustrated in Figure~\ref{fig:xyz}.

This traced origin positions Cluster 2 in the vicinity of the present-day Per OB1 and Cas OB6 associations, which are prominent star-forming complexes in that segment of the Perseus Arm (Figure~\ref{fig:xyz}). While its formation near these major associations is intriguing, Cluster 2's age of $\sim$20 Myr makes it unlikely to be contemporaneous with, or to have triggered, the primary, more recent episodes of star formation observed in either Per OB1 (where triggered stars are $\sim$5-8 Myr old and h/$\chi$ Persei are $\sim$13-14 Myr old; \citealt{2008ApJ...679.1352L, 2002ApJ...576..880S, 2019ApJ...876...65L}) or Cas OB6 (with clusters aged 1-5 Myr; \citealt{2012ApJ...744...87B, 2013A&A...554A...3S, 2014MNRAS.438.1451L}). Its age predates these activities by at least 6-7 Myr. Therefore, Cluster 2 most likely represents an earlier, distinct epoch of star formation within that region of the Perseus Arm, unrelated to the currently dominant, younger populations. While a significantly older age, such as the $\sim$35 Myr upper limit suggested by some NN estimates \citep{2024A&A...686A..42H} (though prone to systematics, see Section~\ref{subsec:age}), could place its origin even deeper within the Perseus Arm assuming its current velocity, our more constrained age of $\sim$20 Myr supports this scenario of predating the currently prominent activity. Any unmodeled acceleration experienced by Cluster 2 could also alter its exact traced origin point.

Definitively linking Cluster 2 to a specific parent molecular cloud or substructure within the Perseus Arm is beyond the scope of this work and would necessitate more comprehensive RV data for all its members and detailed N-body simulations that model its orbit through the Galactic potential, considering the influence of the spiral arms. A comprehensive look at the chemical composition of Cluster 2, could provide additional hints at its origin; however, it is unlikely to offer any definitive conclusions. Many young populations on local scales tend to be chemically homogeneous (e.g., \citealt{2009A&A...501..973D,2011A&A...526A.103D}). There does exist a minor gradient in metal abundances on galactic scales \citep{2023NatAs...7..951L} with slight fluctuations between and within different spiral arm features \citep{2025MNRAS.tmpL..72M}. However, differences in metallicity between the CamOB1 area and the Perseus arm are expected to be less than 0.1 dex (with the Perseus arm being more metal-poor). Fluctuations in metallicity within different regions of the Perseus arm would be even less.

Cluster 2's journey from a different spiral arm to its current location, where it happens to be commingling with two other unrelated clusters, is a remarkable illustration of the dynamic nature of stellar populations within the Galaxy. If a more definitive association with the Per OB1 or Cas OB6 regions could be established in the future, Cluster 2 might offer a valuable glimpse into the conditions and mechanisms of an older generation of star formation in this active part of the Perseus Arm.

\section{Summary} \label{sec:summary}
This work seeks to characterize the location, motion, and origin of three individual clusters in the Cam OB1 association that were previously thought of as one region. Our initial interest in this region of Cam OB1 stemmed from recent en-masse clustering efforts to discover new structures and groupings using Gaia's extensive database of precise stellar parallaxes and proper motion. In the process of en masse searches for open clusters, both \citet{2020AJ....160..279K,2023A&A...673A.114H} listed multiple distinct kinematically coherent groups, all found in a similar location in the sky and at a similar distance.

Using a similar blind clustering in gaia 3D position + 2D proper motion measurements towards the region of interest in Cam OB1, we recover three clusters at the approximate locations noted in prior clustering works. In the process, we refine the membership list, allowing for more extended diffuse members than can be done in previous en masse cluster identifications. We further augment our sample with RVs from SDSS-V where available. We add back in previously classified field sources whose RV falls near the peak of one of the cluster's RV distribution and who shares proper motion, position, and parallax consistent with the cluster.

By solely utilizing astrometry in clustering, we make no limitations on the spectral type. This dramatically enhances the sample size of each cluster (Cluster 1: 469; Cluster 2: 184; and Cluster 3: 140 members) compared to the prior ~98 OB sources for this region as a whole. Previous works analyzing the Cam OB1 region, before Gaia-based astrometry, have predominantly been focused on augmenting the YSOs and OB star inventory. Due to prior uncertainty in astrometry (i.e., membership), YSOs and massive stars with more readily identifiable signatures of youth could provide more direct association with Cam OB1's molecular clouds.

With a larger, more representative sample of all populations, we fit isochrones to each cluster to determine their ages (Cluster 1=15.8 Myr; Cluster 2=20.0 Myr; Cluster 3=10.0 Myr). Utilizing the parallaxes, we note that each cluster currently overlaps in 3D space, not just in projection. Using the subset of sources with RVs, we project each cluster's path back in time to the age and approximate location of its formation.  

Each cluster is shown to have originated from its own distinct region of space. The distances between each cluster at the time of formation prohibit any mutual influence on initial formation. The youngest two clusters (Cluster 1 and 3) appear to have origins from different regions within the Cam OB1 association (approximately ~240pc and 65pc from their respective initial location).

Cluster 2, however, is unique, with age $\sim$5-10Myr older than the other clusters. It possesses a faster 3D motion relative to the other clusters (covering $>$550pc since formation), with its RVs more systematically blue-shifted by $\gtrapprox$21 km/s. The RVs are more consistent with clusters and molecular clouds located in the Perseus arm. The difference in RVs has, in part, led the members of Cluster 2 to be excluded from any previous work analyzing Cam OB1. In lieu of accurate parallaxes, these sources were assumed to be distant Perseus arm objects due to their RVs. 

In fact, tracing back the path of cluster 2 puts its formation location right on the near edge of the Perseus Arm, close to the present-day location of the Cas OB6 and Per OB1 association. The age of Cluster 2 predates the current round of star formation in either Cas OB6 or Per OB1 by $>$6Myr, indicating either an origin distinct from these molecular clouds or a look at an earlier round of star formation within the associations. More intensive modeling of the galactic potential to trace the motion of Cluster 2 and the Perseus arm members will be needed to make any definitive links to its originating molecular cloud. We leave it to future works, following more spectroscopic observations, to further study the three-dimensional kinematic properties of Cluster 2 and its origin. Additional spectroscopy will give RV measurements for further cluster sources and provide a look into the chemical composition of Cluster 2 in comparison to its origin.

We conclude by noting just how wild it is that multiple unrelated populations are currently inhabiting the same region of space, which had zero influence on each other's formation. Additionally, one cluster happens to have its origin tracing back to a completely different spiral arm than its current location in Cam OB1. The chance alignment of all three clusters at the current time just made them particularly prominent to us. There are most likely numerous more systems with visitors like this. This work is an example of the discoveries that can be made by adding RVs (the final dimension) to existing clustering provided by Gaia's precise astrometry.




\begin{acknowledgments}
Funding for the Sloan Digital Sky Survey V has been provided by the Alfred P. Sloan Foundation, the Heising-Simons Foundation, the National Science Foundation, and the Participating Institutions. SDSS acknowledges support and resources from the Center for High-Performance Computing at the University of Utah. SDSS telescopes are located at Apache Point Observatory, funded by the Astrophysical Research Consortium and operated by New Mexico State University, and at Las Campanas Observatory, operated by the Carnegie Institution for Science. The SDSS web site is \url{www.sdss.org}.

SDSS is managed by the Astrophysical Research Consortium for the Participating Institutions of the SDSS Collaboration, including the Carnegie Institution for Science, Chilean National Time Allocation Committee (CNTAC) ratified researchers, Caltech, the Gotham Participation Group, Harvard University, Heidelberg University, The Flatiron Institute, The Johns Hopkins University, L'Ecole polytechnique f\'{e}d\'{e}rale de Lausanne (EPFL), Leibniz-Institut f\"{u}r Astrophysik Potsdam (AIP), Max-Planck-Institut f\"{u}r Astronomie (MPIA Heidelberg), Max-Planck-Institut f\"{u}r Extraterrestrische Physik (MPE), Nanjing University, National Astronomical Observatories of China (NAOC), New Mexico State University, The Ohio State University, Pennsylvania State University, Smithsonian Astrophysical Observatory, Space Telescope Science Institute (STScI), the Stellar Astrophysics Participation Group, Universidad Nacional Aut\'{o}noma de M\'{e}xico, University of Arizona, University of Colorado Boulder, University of Illinois at Urbana-Champaign, University of Toronto, University of Utah, University of Virginia, Yale University, and Yunnan University.
\end{acknowledgments}

\begin{contribution}

All authors contributed to the development and revision of the material presented in this paper.


\end{contribution}

%
\facilities{SDSS, Gaia}

\software{astropy \citep{2013A&A...558A..33A},  
          HDBSCAN \citep{2017JOSS....2..205M}, 
          AURIGA \citep{2020AJ....160..279K}
          }



\bibliographystyle{aasjournalv7}



\end{document}